\documentclass[12pt]{article}
\usepackage{a4wide,graphicx,amsmath,amssymb,cite,isolatin1}
\usepackage{thophys}
\newcommand{\QQ}{Q\overline{Q}}
\newcommand{\mQ}{m_{\!_Q}}
\newcommand{\bs}[1]{\mbox{\boldmath ${#1}$}}
\newcommand{\Mbc}{M_{b\bar{c}}}
\newlength{\Dlrwd}\newlength{\Dlrht}\newsavebox{\Dlrarg}%
\newcommand{\Dlr}{%
  \savebox{\Dlrarg}{\ensuremath{D}}%
  \settowidth{\Dlrwd}{\usebox{\Dlrarg}}%
  \settoheight{\Dlrht}{\usebox{\Dlrarg}}%
  \raisebox{0.2\Dlrht}{\ensuremath{- \tfrac{i}{2}}}
  \raisebox{1.2\Dlrht}{\makebox[0pt][l]{\hspace{0.2\Dlrwd}%
    \scriptsize \ensuremath{\leftrightarrow}}}%
  \usebox{\Dlrarg}}
\newlength{\Dslrwd}\newlength{\Dslrht}\newsavebox{\Dslrarg}%
\newcommand{\Dslr}{%
  \savebox{\Dslrarg}{\ensuremath{\fmslash{D}}}%
  \settowidth{\Dslrwd}{\usebox{\Dslrarg}}%
  \settoheight{\Dslrht}{\usebox{\Dslrarg}}%
  \raisebox{0.2\Dslrht}{\ensuremath{- \tfrac{i}{2}}}
  \raisebox{1.2\Dslrht}{\makebox[0pt][l]{\hspace{0.2\Dslrwd}%
    \scriptsize \ensuremath{\leftrightarrow}}}%
  \usebox{\Dslrarg}}
\newlength{\Dblrwd}\newlength{\Dblrht}\newsavebox{\Dblrarg}%

\newlength{\Dmlrwd}\newlength{\Dmlrht}\newsavebox{\Dmlrarg}%
\newcommand{\Dmlr}{%
  \savebox{\Dmlrarg}{\ensuremath{D}}%
  \settowidth{\Dmlrwd}{\usebox{\Dmlrarg}}%
  \settoheight{\Dmlrht}{\usebox{\Dmlrarg}}%
  \raisebox{0.2\Dmlrht}{\ensuremath{\tfrac{i}{2}}}
  \raisebox{1.2\Dmlrht}{\makebox[0pt][l]{\hspace{0.2\Dmlrwd}%
    \scriptsize \ensuremath{\leftrightarrow}}}%
  \usebox{\Dmlrarg}}
\newlength{\iDlrwd}\newlength{\iDlrht}\newsavebox{\iDlrarg}%
\newcommand{\iDlr}{%
  \savebox{\iDlrarg}{\ensuremath{D}}%
  \settowidth{\iDlrwd}{\usebox{\iDlrarg}}%
  \settoheight{\iDlrht}{\usebox{\iDlrarg}}%
  \ensuremath{i}\raisebox{1.2\iDlrht}{\makebox[0pt][l]{\hspace{0.2\iDlrwd}%
    \scriptsize \ensuremath{\leftrightarrow}}}%
  \usebox{\iDlrarg}}
\newlength{\iDslrwd}\newlength{\iDslrht}\newsavebox{\iDslrarg}%
\newcommand{\iDslr}{%
  \savebox{\iDslrarg}{\ensuremath{\fmslash{D}}}%
  \settowidth{\iDslrwd}{\usebox{\iDslrarg}}%
  \settoheight{\iDslrht}{\usebox{\iDslrarg}}%
  \ensuremath{i}\raisebox{1.2\iDslrht}{\makebox[0pt][l]{\hspace{0.2\iDslrwd}%
    \scriptsize \ensuremath{\leftrightarrow}}}%
  \usebox{\iDslrarg}}
\begin{document}
\begin{titlepage}

\begin{flushright}
FTUV-01-0926\\
TTP01-24\\
hep-ph/0109250\\
\today
\end{flushright}
\vskip 4.0ex
\begin{center}
\boldmath
{\Large \bf Non-perturbative effects in semi-leptonic $B_c$ decays}
\unboldmath
\vskip 6.0ex
{\sc Thomas Mannel}
\vskip 2.0ex
{\em Institut für Theoretische Teilchenphysik, Universität Karlsruhe,\\
D-76128 Karlsruhe, Germany}
\vskip 4.0ex
{\sc Stefan Wolf}
\vskip 2.0ex
{\em Departament de Física Teòrica, Universitat de València\\
Dr.~Moliner 50, E-46100 Burjassot, València, Spain}
\vskip 2.0ex
\end{center}

\begin{abstract}
We discuss the impact of the soft degrees of freedom inside the $B_c$ meson on
its rate in the semi-leptonic decay $B_c \to X \ell \bar{\nu}_\ell$ where $X$
denotes light hadrons below the $D^0$ threshold. In particular we identify
contributions involving soft hadrons which are non-vanishing in the limit of
massless leptons. These contributions become relevant for a measurement of the
purely leptonic $B_c$ decay rate, which due to helicity suppression involves a
factor $m_\ell^2$ and thus is much smaller than the contributions involving
soft hadrons.
\\[2ex]
\end{abstract}
%
\end{titlepage}

\section{Introduction}

Within the standard model of elementary particle physics the $B_c$ is the only
meson which contains a heavy quark-antiquark ($\QQ$) pair of different
flavour. After more then two decades of purely theoretical investigations its
observation by the CDF collaboration in 1998 \cite{Abe:1998wi, Abe:1998fb}
pushed open the gate for experimental exploration. The newest and forthcoming
generation of experiments (Tevatron and LHC) will be able to determine the
$B_c$ properties that are hardly known at present \cite{Groom:2000in}.

Theoretically, the fact that the $B_c$ is a bound state of heavy quarks offers
an interesting possibility for an efficient way of calculating its properties.
While the bound state is dominated by soft scales the heavy quark mass $\mQ$
which is the typical scale for production and decay processes of such mesons is
much larger. Due to this hierarchy of scales it is obvious to use an effective
theory approach. One could treat the $B_c$ within the Heavy Quark Effective
Theory (HQET) at intermediate scales (i.e.~where the charm (anti)quark is
considered to be light) but this does not take into account that $m_c$ can
still be considered a perturbative scale. Assuming that the $c$ quark is also
heavy (i.e.~using that both $m_b, m_c \gg \Lambda_{\rm QCD}$) we may work with
approaches developed for quarkonia-like systems. In this case the $B_c$ stands
somehow in between of charmonia and bottomonia which usually are investigated
within the framework of Non-relativistic Quantumchromodynamics (NRQCD)
\cite{Bodwin:1995jh}.

In this article we concentrate on the semi-leptonic $B_c$ decay $B_c \to X \ell
\bar{\nu}_\ell$ where $X$ denotes only {\it light} hadrons which means hadrons
below the charm threshold. In this kinematic region the $B_c$ has to decay
through electroweak annihilation. The light hadrons thus originate from gluons
or other light degrees of freedom inside the $B_c$ meson which makes this decay
mode particularly interesting, since it tests the light degrees of freedom in a
quarkonia-like system. But also for practical purposes it is of relevance: We
will show that the semi-leptonic decay rate exceeds the leptonic one (at least
for light leptons) even in the limit of extremely soft light hadrons which
could complicate substantially the measurement of the latter.

The total rate of the purely leptonic mode is given by the expression
\begin{equation}
\label{Gamlept}
\Gamma[B_c \to \ell \bar{\nu}_\ell]
= \frac{G_F^2}{8\pi} |V_{cb}|^2 f_{B_c}^2 M^3
\left[ \frac{m_\ell^2}{M^2} \left(1 - \frac{m_\ell^2}{M^2}\right)^2 \right]
\end{equation}
where $G_F$ is the Fermi constant, $V_{cb}$ the CKM matrix element
corresponding to the underlying partonic decay process $b \bar{c} \to \ell
\bar{\nu}_\ell$, and $M$ is the mass of the $B_c$. The $B_c$ decay constant
$f_{B_c}$ is defined via
\begin{equation}
i f_{B_c} p^\mu := \bra{0} \bar{b} \gamma^\mu \gamma^5 c \ket{B_c(p)} \,.
\end{equation}
Obviously the total rate (\ref{Gamlept}) is suppressed by the ratio
$m_\ell^2/M^2$ of the lepton and the $B_c$ mass squared. This is caused by the
helicity flip which is necessary in a back-to-back decay of a spinless particle
in the ground state into a left-handed lepton (right-handed antilepton) and its
right-handed antineutrino (left-handed neutrino).

Contrary to the quark model where the $(b\bar{c})$ systems has to be in a $0^-$
state, NRQCD and related effective approaches allow for higher Fock states,
i.e.~the $(b\bar{c})$ pair also can exist with other quantum numbers than $J^P
= 0^-$ inside the $B_c$. Although these states are suppressed by
non-perturbative factors as $\bar{v}^2$ in NRQCD\footnote{We have chosen
$\bar{v}$ instead of the commonly used $v$ to denote the typical (average)
non-relativistic (anti)quark velocity inside the quarkonia, i.e.~more precisely
the absolute value of its three vector, to avoid confusion with the covariant
four vector $v_\mu$ we need to define the quarkonia velocity.} their
contribution becomes measurable if the $m^2_\ell$ suppression of the leading
Fock state is stronger than the $\bar{v}^2$ suppression of a subleading Fock
state which does not suffer from helicity suppression. However, these
contributions are not purely leptonic anymore since in such a case the soft
degrees of freedom inside the higher Fock states should give rise to some soft
hadronic signal in the final state.

Thus we expect, for instance, semi-leptonic contributions related to the NRQCD
matrix elements $\braket{{\cal O}_1({}^3\!S_1)}$ and $\braket{{\cal
O}_1({}^1\!P_1)}$ where the short distance coefficients arise without any
$m_\ell^2/M^2$ suppression since it is not necessary to flip any helicity if
the decaying $(b\bar{c})$ pair has spin one or is in a $P$ wave state.
Additionally there also should be unsuppressed contributions from higher Fock
states with the $(b\bar{c})$ pair in a ${}^1\!S_0$ state since the soft degrees
of freedom inside the $B_c$ can carry away some angular momentum or spin to
prevent the helicity suppression in the leptonic sector. Thus the corresponding
short distance coefficients should (at least for light leptons) overcome the
enhancement of the leading order matrix element $\braket{{\cal
O}_1({}^1\!S_0)}$ against the higher orders in the non-relativistic
expansion. They also should exceed the short distance coefficients from
radiative leptonic $B_c$ decays where the prize of avoiding helicity
suppression by radiating off an additional photon is paid for by a factor
$\alpha/(4\pi)$.

In order to calculate the lepton energy spectrum in the decay $B_c \to X \ell
\bar{\nu}_\ell$ we apply an effective approach almost equivalent to standard
NRQCD. Even though we use a different notation namely a covariant one borrowed
from HQET our effective Lagrangian matches the NRQCD one. However, we
enlarge the operator basis at dimension-8 by operators related with the
centre-of-mass (cms) momentum of the $\QQ$ pair inside the bound state since we
expect contributions that are not helicity suppressed just from these
operators. According to the standard NRQCD velocity rules the matrix elements
of such operators are suppressed with respect to matrix elements of dimension-8
operators that merely depend on the relative momentum of quark and antiquark.
On the other hand, it is known that these power counting rules are only
applicable for systems which obey the scale hierarchy $\mQ \gg \mQ \bar{v} \gg
\mQ \bar{v}^2 \sim \Lambda_{\rm QCD}$ \cite{Beneke:1997av}.

Fleming et al.~\cite{Fleming:2000ib} proposed to keep the standard NRQCD
velocity rules only for bottomonia while they suggest to switch to a
dimensional power counting in the charmonium sector. The gist of this analysis
is a new weighting of a spin-flip transition that is compatible with data
\cite{Sanchis-Lozano:2001rr} and possibly could explain some of the NRQCD
drawbacks concerning the $J/\psi$ polarization at high transverse momentum and
the endpoint spectrum in radiative quarkonia decays.

Here we do not only discuss the relative weight of a dipole and a spin-flip
transition. Rather the semi-leptonic decay rate may also help to clarify the
power counting of the cms derivatives on the $\QQ$ pair. Our procedure follows
the strategy of Heavy Quarkonia Effective Theory (HQ$\overline{\rm Q}$ET)
\cite{Mannel:1995xh, Mannel:1995xc}. We just eliminate some redundancies and
rotate the operator basis into the familiar one of NRQCD.

In the following we will start with setting up the basics of the effective
approach. We sketch how the Lagrangian is constructed and how $B_c$
annihilation processes are implemented. Furthermore we provide the basis of
four fermion operators up to dimension-8 and discuss the power counting rules.
After applying this framework on the lepton energy spectrum in the decay $B_c
\to X \ell \bar{\nu}_\ell$ we address the structure of its endpoint region. We
also relate our result to the shape function improved one of NRQCD. Finally we
discuss the total semi-leptonic width of the $B_c$ and implied consequences on
measuring its purely leptonic width.

\section{Effective field theory for the $B_c$}
In this section we shall briefly discuss the effective field theory set up as
it is appropriate for the case of a $B_c$ decay. In general, quarkonia are more
complicated than mesons containing only one heavy (anti)quark due to the fact
that several scales are involved, some of which are dynamically generated.

\subsection{Lagrangian and fields}
In the first step one uses the fact that the heavy quark mass is large compared
to any other scale in the problem. Thus one performs a $1/m_Q$ expansion of the
full QCD Lagrangian, which can be done either by integrating out the heavy
degrees of freedom, which are the small components of the heavy quark spinor
\cite{Mannel:1992mc}, or by performing a sequence of Foldy-Wouthuysen
transformations \cite{Korner:1991kf}. In both cases one obtains a $1/m_Q$
expansion of the heavy quark field as well as of the Lagrangian. Although these
expansions look slightly different, the results for the matrix elements are the
same. Up to $1/m_Q^2$ one obtains
\begin{equation}
\label{L_eff}
{\cal L}_v = \bar{h}_v^{(+)} (ivD) h_v^{(+)}
+ \frac{1}{2\mQ} (K_v + M_v) + {\cal O}\left(\frac{1}{\mQ^2}\right)
\end{equation}
with
\begin{subequations}
\begin{align}
K_v &= \bar{h}_v^{(+)} iD^\mu (g_{\mu\nu} - v_\mu v_\nu) iD^\nu h_v^{(+)} \,,
\\
M_v &= \frac{ig_s}{2} \bar{h}_v^{(+)} (-i\sigma^{\mu\nu}) G_{\mu\nu} h_v^{(+)}
\end{align}
\end{subequations}
where $v_\mu = P_\mu/M$ is the four velocity of the heavy meson and $h_v^{(+)}$
the field operator of a heavy quark in the limit $\mQ \to \infty$. The terms
$K_v$ and $M_v$ correspond to the kinetic energy and chromomagnetic moment,
respectively, where the gluonic field strength tensor $G_{\mu\nu}$ is defined
by the relation
\begin{equation}
i g_s G_{\mu\nu} := [iD_\mu, iD_\nu] \,.
\end{equation}

Likewise, we obtain the corresponding relations for an antiquark in the
infinite mass limit by reversing the sign of the velocity $v \to -v$.

In order to describe a quarkonium-like system one wants to treat both the quark
and the antiquark in the infinite mass limit, assuming that both are sitting
in a bound state moving with a velocity $v$ and that both heavy constituents
have small residual momentum. Thus one would start with the Lagrangian
\begin{equation}
\label{L_stat}
{\cal L}_{\rm stat.}
= \bar{h}_v^{(+)} (ivD) h_v^{(+)} - \bar{h}_{v'}^{(-)} (iv'D) h_{v'}^{(-)}
\end{equation}
and add the corresponding pieces of the Lagrangian for the light degrees of
freedom.

However, it is known that this simple ansatz misses an important piece of
physics, which is that the dynamics responsible for the binding generates
additional scales. These scales are the size and the binding energy of the
bound system and are related to the relative (three-)velocity $v_{\rm rel}$
which can be written as
\begin{equation}
v_{\rm rel} = \sqrt{(vv')^2 - 1} \,.
\end{equation}

Clearly in a quarkonium we have $v_{\rm rel} \ll 1$ and in the limit $v \to v'$
the Lagrangian (\ref{L_stat}) exhibits certain pathologies indicating that
bound states cannot be described in the static limit. As a example one may
consider the matrix element $\bra{H_{\rm stat.}} \bar{h}_v^{(+)} \Gamma
h_{v'}^{(-)} \ket{0}$ in the static limit; computation of the anomalous
dimension reveals an imaginary part in the anomalous dimension of the
respective current $\bar{h}_v^{(+)} \Gamma h_{v'}^{(-)}$ as $v' \to v$
\cite{Grinstein:1991kw}.  The divergence in the limit $v' \to v$ leads to a
phase factor in the solution of the renormalization group equation which is
related to the coulombic part of the one-gluon exchange \cite{Kilian:1993tj,
Kilian:1993nk}. This reflects the possibility of heavy bound states.

This problem can be fixed by including the kinetic term into the leading order
Lagrangian, i.e.~one switches to a non-relativistic description. The Lagrangian
then reads
\begin{equation}
\label{L_0}
{\cal L}_0
= \bar{h}_v^{(+)}
	\left( ivD - \frac{\left(iD_\perp\right)^2}{2\mQ} \right) h_v^{(+)}
- \bar{h}_v^{(-)}
	\left( ivD + \frac{\left(iD_\perp\right)^2}{2\mQ} \right) h_v^{(-)}
\end{equation}
where the component $iD^\mu_\perp$ orthogonal the $v_\mu$ is given by
\begin{equation}
iD^\mu_\perp := (g^{\mu\nu} - v^\mu v^\nu) iD_\nu \,.
\end{equation}
The form of the leading order term evidently shows that is impossible to
operate with a purely dimensional power counting in this case, since the
equation of motion is not homogeneous in $1/\mQ$:
\begin{equation}
\label{eom}
\left(ivD\right) h_v^{(\pm)}
= \pm \frac{\left(iD_\perp\right)^2}{2\mQ} h_v^{(\pm)} \,.
\end{equation}
Usually it is argued that this corresponds to an expansion in the relative
velocity $\bar{v}$ which is assumed to be small. In fact, the non-relativistic
approximation (\ref{L_0}) can generate bound states of (inverse) size $\mQ
\bar{v}$ and of binding energy $\mQ \bar{v}^2$ which introduce two more,
dynamically generated scales into the problem. This fact makes the 
construction of an effective field theory approach more complicated.

We also note that the chromomagnetic moment $M_v$ is taken to be higher order
compared to the kinetic energy $K_v$ although it is of the same mass
dimension. This means that the antisymmetric combination of two covariant
derivatives acting on the field $h_v^{(\pm)}$ is suppressed with respect to the
symmetric combination:
\begin{equation}
\label{iDiD}
 \left[iD^\mu, iD^\nu\right]  h_v^{(\pm)} \quad \ll \quad
\left\{iD^\mu, iD^\nu\right\} h_v^{(\pm)} \,.
\end{equation}
The standard rules for power counting in non-relativistic system can be found
in \cite{Lepage:1992tx}, but we shall return to this question at the end of
this section.

The static as well as the non-relativistic limit of QCD has (compared to full
QCD) an additional spin symmetry which is evident from the Lagrangian
(\ref{L_0}). In the case of a quarkonium-like system this symmetry predicts
quadruplets which are degenerate up to terms of order $\bar{v}^2$. For the
ground state quarkonia this quadruplet consists of the $0^-$ states and the
three polarizations of the $1^-$ state. For excited quarkonia this quadruplet
structure is more complicated and involves a possible orbital angular momentum.

Finally one may formulate NRQCD in the rest frame of the quarkonium in which $v
= (1, \bs{0})$ which transforms (\ref{L_0}) into the usual NRQCD Lagrangian.
Rewriting the (relativistic) four-dimensional fields $h_v^{(\pm)}$ in terms of
the (non-relativistic) two-dimensional spinors $\psi$ and $\chi$ for quark and
antiquark, respectively,
\begin{equation}
\label{4to2}
h_v^{(+)} = \left( \begin{array}{c} \psi \\ 0 \end{array} \right)
\,, \qquad
h_v^{(-)} = \left( \begin{array}{c} 0 \\ \chi \end{array} \right)
\end{equation}
one actually gets the NRQCD Lagrangian
\begin{equation}
{\cal L}_0 = \psi^\dag \left( iD_t + \frac{(i\bs{D})^2}{2\mQ} \right) \psi
           + \chi^\dag \left( iD_t - \frac{(i\bs{D})^2}{2\mQ} \right) \chi \,.
\end{equation}

\subsection{Operator Product Expansion for semileptonic $B_c$ decays}

The main contribution to semileptonic $B_c$ decays comes from the part of the
effective Hamiltonian mediating $b \to c $ transitions
\begin{equation}
H_{\rm eff} = \frac{4 G_F}{\sqrt{2}} V_{cb}(\bar{b}_L \gamma_\mu c_L)
                                     (\bar{\nu}_{\ell,L} \gamma^\mu \ell_L)
\end{equation}
where the subscript $L$ denotes the left-handed component of the the
corresponding fermion.

In order to consider the inclusive case we write the rate as
\begin{equation}
d\Gamma
= \frac{8 G_F^2 |V_{cb}|^2}{M} \int \!\! \frac{d^4q}{(2\pi)^4}
\int \!\! d^4x \, e^{-iqx} \bra{B_c} (\bar{b}_L(x) \gamma_\mu c_L(x))
				     (\bar{c}_L(0) \gamma_\nu b_L(0)) \ket{B_c}
I^{\mu\nu}(q) 
\end{equation}
where $I^{\mu\nu}(q)$ denotes the leptonic part.

We perform an operator product expansion (OPE) for the product of the hadronic
currents making use of the fact that both $m_c$ and $m_b$ are large scales,
which can be removed from the matrix element by redefining the phase of the
heavy fields. We consider the decay of a $B_c^-$ which consists of a anti-$c$
and a $b$ quark; therefore we define
\begin{equation}
b(x) =: e^{-i m_b v x} B_v(x) \quad and \quad c(x) =: e^{i m_c v x} C_v(x)
\end{equation}
where $B_v$ and $C_v$ eventually become the nonrelativistic fields considered
in the last paragraph, and $v$ is the four velocity of the $B_c$ meson. In
terms of these fields we have for the hadronic tensor
\begin{equation}
H_{\mu \nu}
= \int \!\! d^4x \, \exp\{-i (q - \Mbc v) x\}
\bra{B_c(v)} (\bar{B}_{v, L}(x) \gamma_\mu C_{v, L}(x))
             (\bar{C}_{v, L}(0) \gamma_\nu B_{v, L}(0)) \ket{B_c(v)}
\end{equation}
where
\begin{equation}
\Mbc = m_b + m_c
\end{equation}
is the sum of the quark masses.

We perform an OPE assuming that $(q - \Mbc v)^2$ and $qv - \Mbc$ are of order
$\Mbc^2$ and $\Mbc$, respectively. The differential rate takes the form
\begin{equation}
\label{expansion}
d\Gamma = \sum_n {\cal C}_n(\mu) \braket{{\cal O}[n]} \Big|_\mu
\end{equation}
where ${\cal C}_n$ are short distance coefficients which are calculated in
perturbation theory and $\langle {\cal O}[n] \rangle$ are forward
matrix elements of local operators ${\cal O}[n]$ with non-relativistic
$B_c$ states. The parameter $\mu$ is the renormalization scale; the
dependence of the coefficients on $\mu$ is canceled by the corresponding
dependence of the matrix elements, such that the total rate is independent of
$\mu$. The sum runs over an infinite set of operators which are
of increasing mass dimension; this dimension is compensated by the
coefficients ${\cal C}_n$ which contain only the perturbative and large mass
scale $M_{b\bar{c}}$, while the matrix elements
$\langle {\cal O}[n] \rangle$ do not depend on this mass scales any more;
they depend only on a non-perturbative small mass scale $\bar{\Lambda}$
corresponding e.g.~to the binding energy of a coulombic system. 
Thus higher dimensional operators are expected to
be suppressed by powers of $\bar{\Lambda} / M_{b \bar{c}}$ and thus
the series in (\ref{expansion}) may be truncated and we shall consider
only the first nontrivial terms. 

In leading order we have two operators of dimension-6:
\begin{subequations}
\label{1x1}
\begin{align}
{\cal O}_{(1 \times 1)}[{}^1\!S_0^{(C)}]
&:= - \left( \bar{h}_v^{(+)} \, \gamma_5 C \, h_v^{(-)} \right)
      \left( \bar{h}_v^{(-)} \, \gamma_5 C \, h_v^{(+)} \right) \,,
\\
{\cal O}_{(1 \times 1)}[{}^3\!S_1^{(C)}]
&:= - \left( \bar{h}_v^{(+)} \, \gamma_\mu^\perp C \, h_v^{(-)} \right)
      \left( \bar{h}_v^{(-)} \, \gamma_\perp^\mu C \, h_v^{(+)} \right) \,.
\end{align}
\end{subequations}
Here we have introduced the spectral notation $n = {}^{2S+1}\!L_J^{(C)}$ where
the upper index $C = 1,8$ denote the the colour states that are indicated by $C
= 1$ and $C = T^a$ for singlet and octet states, respectively, on the right
hand side of equations (\ref{1x1}). Using (\ref{4to2}) we recognize that the
leading order four fermion operators again coincide with the one of NRQCD
\cite{Bodwin:1995jh}:
\begin{equation}
{\cal O}_{({\rm 1} \times {\rm 1})}[{}^1\!S_0^{(C)}]
\equiv {\cal O}_C({}^1\!S_0) \,,
\qquad
{\cal O}_{({\rm 1} \times {\rm 1})}[{}^3\!S_1^{(C)}]
\equiv {\cal O}_C({}^3\!S_1) \,.
\end{equation}

In subleading order our set of operators is more general than the standard
NRQCD basis, since we also include contributions related to the motion of the
$\QQ$ pair, more precisely its centre of mass, inside the quarkonium. It is
known that these contributions are important in certain kinematical regions,
e.g.~in the endpoint of photon spectra in radiative quarkonia decays
\cite{Mannel:1997uk, Rothstein:1997ac}, even though they are missed by naive
use of NRQCD. To disentangle the relative motion of the quark and the
antiquark from the motion of its centre of mass we decompose their residual
momentum inside the meson according to
\begin{itemize}
\item Residual cms~momentum (RCM):
\begin{equation}
\label{RCM}
iD_\mu \left( \bar{Q}_v^{(\pm)} \, \Gamma C \, Q_v^{(\mp)} \right)
:= \bar{Q}_v^{(\pm)} \, \Gamma C (iD_\mu) \, Q_v^{(\mp)}
+ \bar{Q}_v^{(\pm)} \,(\stackrel{\longleftarrow}{iD_\mu}) \Gamma C\,Q_v^{(\mp)}
\end{equation}
\item Residual relative momentum (RRM):
\begin{equation}
\label{RRM}
\bar{Q}_v^{(\pm)} \, \Gamma \left( \Dmlr_\mu \right) C \, Q_v^{(\mp)}
:= \frac{1}{2} \left[ \bar{Q}_v^{(\pm)} \, \Gamma C (iD_\mu) \, Q_v^{(\mp)}
- \bar{Q}_v^{(\pm)} \,(\stackrel{\longleftarrow}{iD_\mu}) \Gamma C\,Q_v^{(\mp)}
\right]
\end{equation}
\end{itemize}
Besides we introduce $T^{(\mu\nu)} := (T^{\mu \nu} + T^{\nu\mu})/2 - g^{\mu\nu}
T_\lambda^{\,\,\lambda}/4$ to denote the symmetric traceless part of a
tensor. In this way we can distinguish three types of dimension-8 operators:
\\
{\bf RCM $\times$ RCM:}
\begin{subequations}
\begin{align}
{\cal O}_{({\rm C} \times {\rm C})}[{}^1\!S_0^{(C)}]
&:= - \left[ iD_\mu^\perp \left(
         \bar{h}_v^{(+)} \, \gamma_5 C \, h_v^{(-)} \right) \right]
    \left[ iD_\perp^\mu \left(
         \bar{h}_v^{(-)} \, \gamma_5 C \, h_v^{(+)} \right) \right]
\\
{\cal O}_{({\rm C} \times {\rm C})}[{}^3\!S_1^{(C)}]
&:= - \left[ iD_\mu^\perp \left(
         \bar{h}_v^{(+)} \, \gamma_\nu^\perp C \, h_v^{(-)} \right) \right]
    \left[ iD_\perp^\mu \left(
         \bar{h}_v^{(-)} \, \gamma_\perp^\nu C \, h_v^{(+)} \right) \right]
\\
{\cal P}_{({\rm C} \times {\rm C})}[{}^3\!S_1^{(C)}]
&:= - \left[ iD_{(\mu}^\perp \left(
         \bar{h}_v^{(+)} \, \gamma_\perp^\mu C \, h_v^{(-)} \right) \right]
    \left[ iD_{\nu)}^\perp \left(
         \bar{h}_v^{(-)} \, \gamma_\perp^\nu C \, h_v^{(+)} \right) \right]
\end{align}
\end{subequations}

{\bf RCM $\times$ RRM:}
\begin{subequations}
\begin{align}
{\cal O}_{({\rm C} \times {\rm R})}[{}^1\!S_0^{(C)}]
&:= - \frac{1}{2} \left\{
    \left[ \rule{0em}{2.5ex} iD_\mu^\perp \left(
         \bar{h}_v^{(+)} \, \gamma_5 C \, h_v^{(-)} \right) \right]
    \left[
         \bar{h}_v^{(-)} \, \gamma_5
                            \left( \Dlr_\perp^\mu \right) C \, h_v^{(+)}
    \right]
  + {\rm h.c.} \right\}
\\
{\cal O}_{({\rm C} \times {\rm R})}[{}^3\!S_1^{(C)}]
&:= - \frac{1}{2} \left\{
    \left[ \rule{0em}{2.5ex} iD_\mu^\perp \left(
         \bar{h}_v^{(+)} \, \gamma_\nu^\perp C \, h_v^{(-)} \right) \right]
    \left[ \bar{h}_v^{(-)} \, \gamma_\perp^\nu
                            \left( \Dlr_\perp^\mu \right) C \, h_v^{(+)}
    \right]
  + {\rm h.c.} \right\}
\\
{\cal P}_{({\rm C} \times {\rm R})}[{}^3\!S_1^{(C)}]
&:= - \frac{1}{2} \left\{
    \left[ \rule{0em}{2.5ex} iD_{(\mu}^\perp \left(
         \bar{h}_v^{(+)} \, \gamma_\perp^\mu C \, h_v^{(-)} \right) \right]
    \left[ \bar{h}_v^{(-)} \, \gamma_\perp^\nu
                            \left( \Dlr_{\nu)}^\perp \right) C \, h_v^{(+)}
    \right]
  + {\rm h.c.} \right\}
\end{align}
\end{subequations}

{\bf RRM $\times$ RRM:}
\begin{subequations}
\begin{align}
{\cal O}_{({\rm R} \times {\rm R})}[{}^1\!P_1^{(C)}]
&:= \left[
       \bar{h}_v^{(+)} \, \gamma_5 \left( \Dlr_\mu^\perp \right) C \, h_v^{(-)}
    \right]
    \left[
       \bar{h}_v^{(-)} \, \gamma_5 \left( \Dlr_\perp^\mu \right) C \, h_v^{(+)}
    \right]
\\
{\cal O}_{({\rm R} \times {\rm R})}[{}^3\!P_0^{(C)}]
&:= \frac{1}{3} \left[
       \bar{h}_v^{(+)} \left( \Dslr^\perp \right) C \, h_v^{(-)}
    \right]
    \left[
       \bar{h}_v^{(-)} \left( \Dslr^\perp \right) \! C \, h_v^{(+)}
    \right]
\\
{\cal O}_{({\rm R} \times {\rm R})}[{}^3\!P_1^{(C)}]
&:= \frac{1}{2} \left[ i \epsilon_\mu^{\;\;\alpha\beta\gamma} v_\gamma \,
       \bar{h}_v^{(+)} \, \gamma_\alpha^\perp
                          \left( \Dlr_\beta^\perp \right) C \, h_v^{(-)}
    \right]
    \left[ i \epsilon^{\mu{\alpha'}{\beta'}{\gamma'}} v_{\gamma'} \,
       \bar{h}_v^{(-)} \, \gamma_{\alpha'}^\perp
                          \left( \Dlr_{\beta'}^\perp \right) C \, h_v^{(+)}
    \right]
\\
{\cal O}_{({\rm R} \times {\rm R})}[{}^3\!P_2^{(C)}]
&:= \left[
       \bar{h}_v^{(+)} \, \gamma_{(\mu}^\perp
                          \left( \Dlr_{\nu)}^\perp \right) C \, h_v^{(-)}
    \right]
    \left[
       \bar{h}_v^{(-)} \, \gamma_\perp^{(\mu}
                          \left( \Dlr_\perp^{\nu)} \right) C \, h_v^{(+)}
    \right]
\end{align}
\end{subequations}
\begin{subequations}
\begin{align}
{\cal P}_{({\rm R} \times {\rm R})}[{}^1\!S_0^{(C)}]
&:= \frac{1}{2} \left\{
    \left[ \rule{0em}{3ex}
       \bar{h}_v^{(+)} \, \gamma_5 C \, h_v^{(-)}
    \right]
    \left[
       \bar{h}_v^{(-)} \, \gamma_5 \left( \Dlr^\perp \right)^2 C \, h_v^{(+)}
    \right]
  + {\rm h.c.} \right\}
\\
{\cal P}_{({\rm R} \times {\rm R})}[{}^3\!S_1^{(C)}]
&:= \frac{1}{2} \left\{
    \left[ \rule{0em}{3ex}
       \bar{h}_v^{(+)} \, \gamma_\mu^\perp C \, h_v^{(-)}
    \right]
    \left[
       \bar{h}_v^{(-)} \, \gamma_\perp^\mu
                          \left( \Dlr^\perp \right)^2 C \, h_v^{(+)}
    \right]
  + {\rm h.c.} \right\}
\\
{\cal P}_{({\rm R} \times {\rm R})}[{}^3\!S_1^{(C)}, {}^3\!D_1^{(C)}]
&:= \frac{1}{2} \left\{
    \left[ \rule{0em}{2.5ex}
       \bar{h}_v^{(+)} \, \gamma_\mu^\perp C \, h_v^{(-)}
    \right]
    \left[
       \bar{h}_v^{(-)} \, \gamma_\nu^\perp \left( \Dlr_\perp^{(\mu} \right)
                          \left( \Dlr_\perp^{\nu)} \right) C \, h_v^{(+)}
    \right]
  + {\rm h.c.} \right\}
\end{align}
\end{subequations}

The main difference to the operator basis of NRQCD is the inclusion of
operators of the type RCM $\times$ RCM and RCM $\times$ RRM. Dealing with the
cms momentum of the $\QQ$ pair inside the quarkonium they were neglected in the
original work \cite{Bodwin:1995jh} since they do not yield contributions from
the dominant Fock state $\ket{\QQ}$. Nevertheless it was shown that they play
an important role in quarkonia decay \cite{Rothstein:1997ac} and production
\cite{Beneke:1997qw} processes as long as the observable is not inclusive
enough. More precisely these operators are responsible for the shift from the
partonic to the hadronic endpoint in energy spectra caused by the mass
difference $M - 2\mQ \sim \mQ \bar{v}^2$. After resumming the leading cms
operators to so-called shape functions the NRQCD prediction is extendible up to
higher energy values. Numerical analyses of this effect using a model
consistent with the shape function formalism indicate on the one hand a natural
solution of the HERA problem in the $J/\psi$ photoproduction channel
\cite{Beneke:2000gq, Wolf:2001vk} but generate on the other hand some new
obscurities in the radiative $\Upsilon$ decay \cite{Wolf:2001pm}.

The operators or the RRM $\times$ RRM sector common with NRQCD are related to
its dimension-8 operators in the following way (cf.~eq.~(\ref{4to2})):
\begin{equation}
{\cal O}_{({\rm R} \times {\rm R})}[{}^{2S+1}\!L_J^{(C)}]
= {\cal O}_C({}^{2S+1}\!L_J) \,,
\qquad
{\cal P}_{({\rm R} \times {\rm R})}[{}^{2S+1}\!L_J^{(C)}]
= {\cal P}_C({}^{2S+1}\!L_J) \,.
\end{equation}

\subsection{Power counting}

\begin{figure}[H]
  \begin{center}
    \leavevmode
    \includegraphics[]{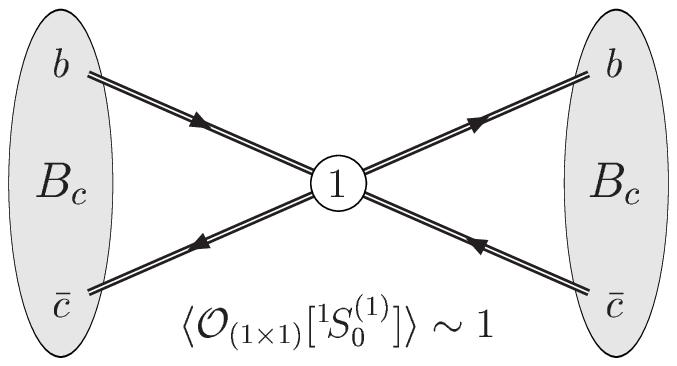}
    \quad
    \includegraphics[]{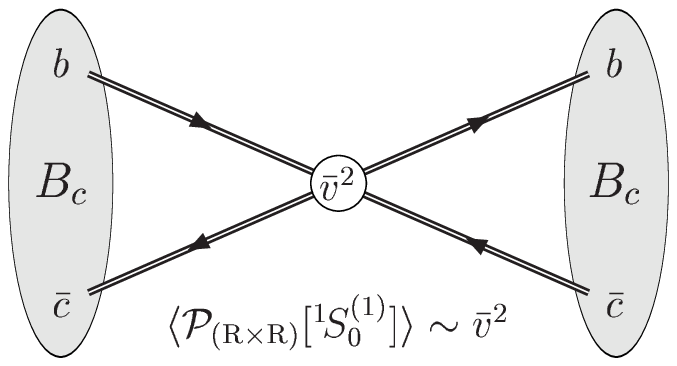}
    \\[1ex]
    \includegraphics[]{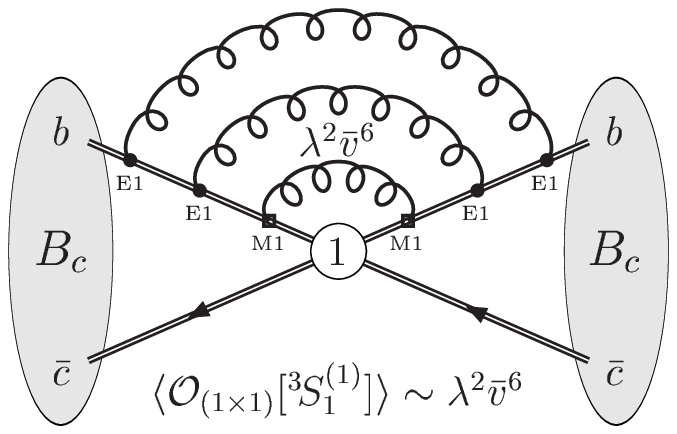}
    \quad
    \includegraphics[]{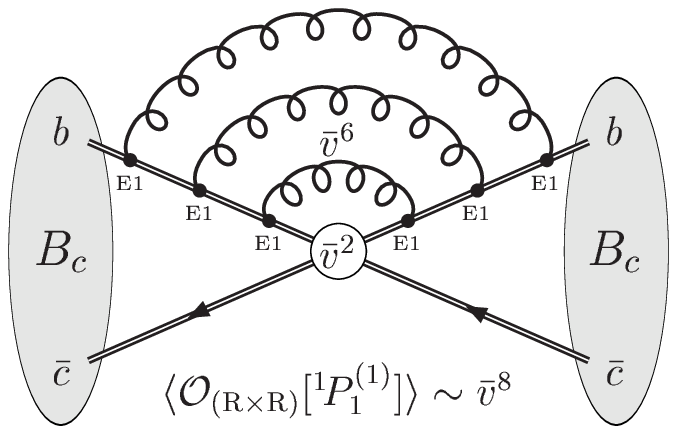}
    \\[1ex]
    \includegraphics[]{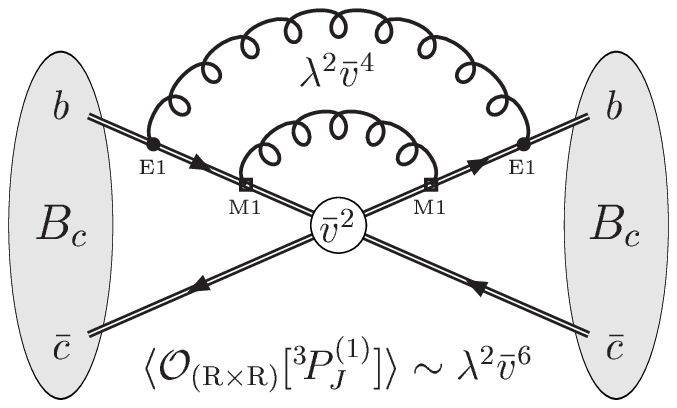}
    \quad
    \includegraphics[]{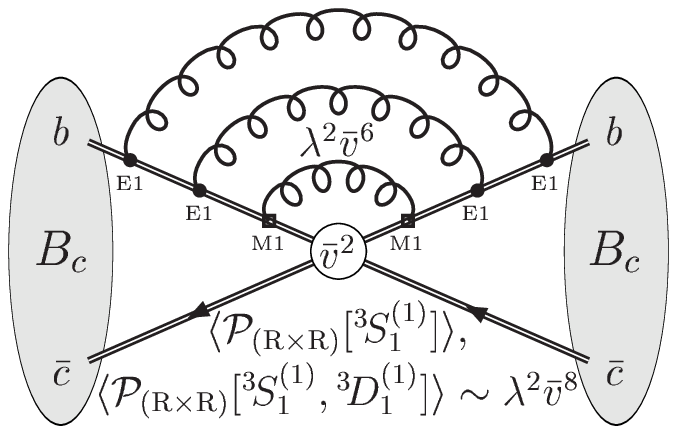}
    \\[1ex]
    \includegraphics[]{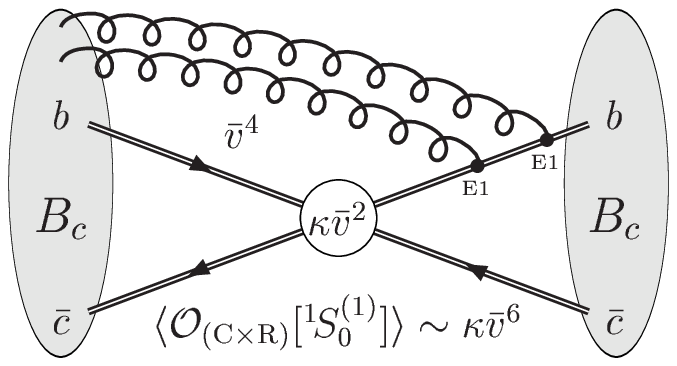}
    \quad
    \includegraphics[]{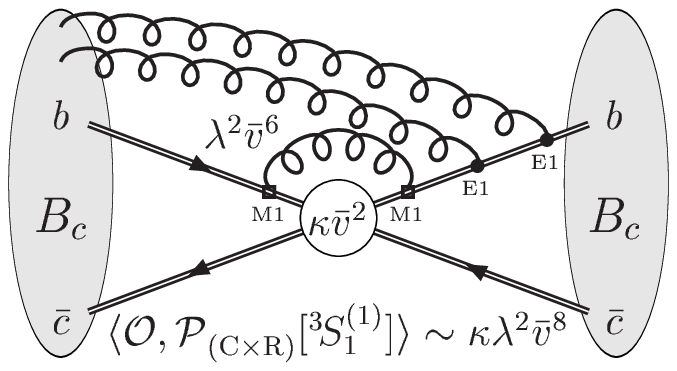}
    \\[1ex]
    \includegraphics[]{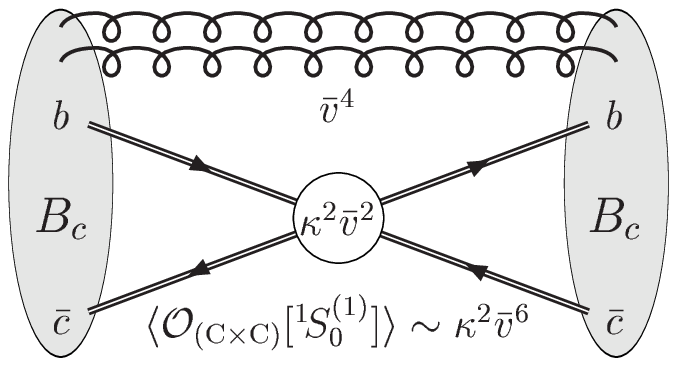}
    \quad
    \includegraphics[]{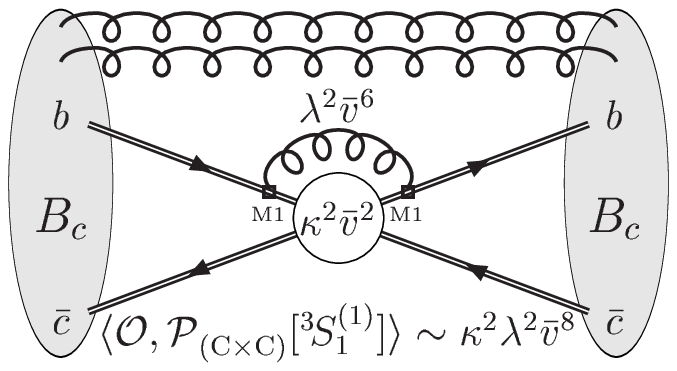}
  \end{center}
    \vspace{-3.5ex}
  \caption{Scaling rules for $B_c$ matrix elements of operators up to
	dimension-8. The operator power counting is denoted inside the central
	circle, the Fock state suppression above.}
  \label{fig1}
\end{figure}

Instead of applying strictly the standard NRQCD velocity scaling rules we
propose to use a slightly generalized power counting scheme. In particular we
want to direct the readers attention on contributions related to the cms
movement of the $\QQ$ pair inside the meson. These contributions potentially
exceed the prediction by standard NRQCD power counting, which we will
critically recapitulate first.

To determine the suppression factor of NRQCD matrix elements of four-fermion
operators we have to consider both the scaling of the operator itself and
additional orders of $\bar{v}$ coming from the Fock decomposition of the
hadronic state:
\begin{equation}
\label{Fock}
\ket{H} = \sum_n \left( c^H_0[n] \ket{\QQ[n]}
	+ c^H_1[n] \ket{\QQ[n]g} + c^H_2[n] \ket{\QQ[n]gg} + \ldots \right) .
\end{equation}
While the standard velocity rules for the operators are straightforward the
assignment of suppression factors to the coefficients $c_i[n]$ in (\ref{Fock})
is quite complicated since it depends on the non-perturbative dynamics inside
the bound state. In the original NRQCD paper \cite{Bodwin:1995jh} quarkonia
states are assumed to obey the scale hierarchy $\mQ \gg \mQ \bar{v} \gg \mQ
\bar{v}^2 \sim \Lambda_{\rm QCD}$. In other words their binding is assumed to
be coulombic which enables the use of NRQED velocity rules of
\cite{Lepage:1992tx}. In particular this implies the validity of multipole
expansion for calculating the coefficients, i.e.~a spin-flip transition (M1) is
suppressed by $\bar{v}^2$ in contrast to a dipole transition (E1) which only
accounts for $\bar{v}$.

Beneke mentioned that this is no longer true if the scale hierarchy is $\mQ \gg
\mQ \bar{v} \sim \Lambda_{\rm QCD} \gg \mQ \bar{v}^2$ \cite{Beneke:1997av}. He
proposed to consider M1 transitions as $\lambda \bar{v}$ where $\lambda =
\bar{v}$ for coulombic systems while $\lambda \sim {\cal O}(1)$ for $\mQ
\bar{v} \sim \Lambda_{\rm QCD}$. A recent analysis \cite{Fleming:2000ib}
elaborates this feature to claim that the standard NRQCD velocity rules are
only applicable in the bottomonium sector while it suggests a HQET-like power
counting scheme in $\Lambda_{\rm QCD}/\mQ$ for charmonia.

We take a further step towards generalizing the counting rules questioning the
scaling of a cms derivative inside a NRQCD operator. According to
\begin{equation}
|\bs{p}_{\QQ}| = |\bs{p}_{\rm soft}| = E_{\rm soft}
= \frac{M^2 - 4m_Q^2}{2M} = \frac{M + 2m_Q}{2M} (M - 2m_Q)
\end{equation}
the total momentum $\bs{p}_{\QQ}$ of the $\QQ$ pair inside a quarkonium is
connected with its binding energy $M - 2m_Q$. Therefore in the standard
formalism the cms derivative on a quark-antiquark bilinear accounts as
$\bar{v}^2$. However, if the binding of the quarkonia-like system is not
coulombic anymore the binding energy no longer should be purely coulombic
either. Rather there also should be non-negligible contributions of order
$\Lambda_{\rm QCD}/\mQ \gg \bar{v}^2$. Hence we assume a cms derivative to be
of order $\kappa \bar{v}$ keeping in mind that the standard NRQCD power
counting is restored by $\kappa = \bar{v}$.

With this set of rules the $B_c$ decay matrix elements of the operators up to
dimension-8 scale as indicated in Fig.~\ref{fig1}. Note that colour selection
rules forbid the existence of solely one gluon in the higher Fock state if we
restrict the perturbative part to leading order $\alpha_s$. To estimate the
leading contributions according to the non-relativistic expansion we have to
know if the quarkonia-like $B_c$ behaves rather as bottomonia or as
charmonia. In the first case ($\kappa, \lambda \to \bar{v}$) the first five
configurations of Fig.~\ref{fig1} would contribute up to order ${\cal
O}(\bar{v}^8)$ while for $\kappa, \lambda \to 1$ the main matrix elements are
the ones of the first row and the first column in Fig.~\ref{fig1}.

\section{Lepton energy spectrum}

Using these assumptions we shall now calculate the lepton energy spectrum
$d\Gamma/dz$ in the semi-leptonic decay $B_c \to X \ell \bar{\nu}_\ell$ where
$X$ denote only light hadrons.
\begin{equation}
z := \frac{2 E_\ell}{\Mbc}
\end{equation}
is the lepton energy $E_\ell$ in the $B_c$ rest frame normalized on half of the
sum of the heavy quark masses.

Note that $\Mbc$ is slightly smaller than the mass $M$ of the $B_c$. The mass
difference $M - \Mbc$ just count for the binding energy of the ($b\bar{c}$)
system inside the $B_c$ meson which is of the order $\bar{v}^2$ within usual
NRQCD. However, this has to be taken into account to get the correct endpoint
of the energy spectrum \cite{Rothstein:1997ac}. We will return to this point
later again.

The underlying partonic process is the electroweak decay $b + \bar{c} \to \ell
\bar{\nu}_\ell$. Thus we get
\begin{equation}
\frac{d\Gamma}{dz}
= \frac{1}{M} \left( \frac{4 G_F}{\sqrt{2}} \right)^2 |V_{cb}|^2
\int \!\! d^4x \, \frac{d^4q}{(2\pi)^4} \, e^{- i x [q - \Mbc v]}
\, I_{\mu\nu}(q) \, \bra{B_c} J_v^\mu(x) \bar{J}_v^\nu(0) \ket{B_c} \,.
\end{equation}
The $B_c$ rest frame is characterized by $v_\mu = (1,\bs{0})$. There the tensor
$I_{\mu\nu}$ solely depends on the momentum $q$ that flow through the effective
vertex
\begin{equation}
I_{\mu\nu}(q)
= \int \!\! \widetilde{dp}_\ell \widetilde{dp}_{\bar{\nu}_\ell} \,
(2\pi)^4 \delta^4(q - p_\ell - p_{\bar{\nu}_\ell}) \, L_{\mu\nu} \,
\delta(z - \frac{2 p^0_\ell}{\Mbc})
\end{equation}
which means simultaneously that only the symmetric terms of the leptonic tensor
\begin{equation}
\label{lepten}
L^{\mu\nu}
= 2 \left(p^\mu_\ell p^\nu_{\bar{\nu}_\ell} + p^\nu_\ell p^\mu_{\bar{\nu}_\ell}
+ g^{\mu\nu} (p_\ell \cdot p_{\bar{\nu}_\ell})
- i \epsilon^{\mu\nu}_{\;\;\;\rho\sigma} p^\rho_\ell p^\sigma_{\bar{\nu}_\ell}
\right)
\end{equation}
survive. Let us turn to the hadronic matrix element now. Since we have
factorized off their main phase space dependence the fields $B_v(x)$ and
$C_v(x)$ in the left-handed current
\begin{equation}
J_v^\mu(x) = \bar{B}_v(x) \frac{\gamma^\mu (1 - \gamma_5)}{2} C_v(x)
\end{equation}
only depend on residual momenta. Hence we can expand the hadronic matrix
element in the coordinate space around $x = 0$ after we have ensured
hermiticity by symmetrizing the matrix element previously:
\begin{equation}
\bra{B_c} J_v^\mu(x) \bar{J}_v^\nu(0) \ket{B_c}
\to \bra{B_c} \tfrac{1}{2} [J_v^\mu(x) \bar{J}_v^\nu(0)
                          + J_v^\mu(0) \bar{J}_v^\nu(-x)] \ket{B_c}
=: R_v^{\mu\nu} \,.
\end{equation}
Then the expansion in $x$ correspond to an expansion in the residual cms
momentum of the ($b\bar{c}$) pair inside the $B_c$. Up to second order we 
get 
\begin{equation}
R_v^{\mu\nu}(x)
= R_v^{\mu\nu}(0)
+ \left. i\partial^\alpha R_v^{\mu\nu}(x) \right|_{x=0} (- i x_\alpha)
+ \frac{1}{2}
  \left. i\partial^\alpha i\partial^\beta R_v^{\mu\nu}(x) \right|_{x=0}
                                          (- i x_\alpha) (- i x_\beta) + \ldots
\end{equation}
where the currents on their parts are also expansions parameterizing the
residual relative momentum of the ($b\bar{c}$) system. It is given in
eq.~(\ref{bilin}) of the Appendix.

The $x$ dependence in the phase space integrals of the form
\begin{equation}
\int \!\! d^4x \, \frac{d^4q}{(2\pi)^4} \,
\left\{ 1, (- i x_{\alpha}), (- i x_{\alpha})(- i x_{\beta}) \right\}
 e^{- i x [q - \Mbc v]} \, I_{\mu\nu}(q)
\end{equation}
is regarded by expressing the factor $- i x_\alpha$ as derivative
$\partial/\partial q^\alpha$ on the exponential and subsequent integration by
parts. At this stage arise higher derivative of delta functions, and 
the result for the spectrum is of the structure
\begin{equation}
\label{dGdz}
\frac{d\Gamma}{dz} =  \Gamma_{\!0} \left\{
  {\cal A}   \delta(1 + \varepsilon - z) + \frac{1}{(2m)^2} \left[
  {\cal B}_0 \delta(1 + \varepsilon - z)
+ {\cal B}_1 \delta'(1 + \varepsilon - z)
+ {\cal B}_2 \delta''(1 + \varepsilon - z) \right] \right\} .
\end{equation}

The fact that the differential decay rate is proportional to localized
distributions (i.e.~$\delta$ functions and its derivatives) reflects the two
body structure of the decay at (partonic) tree level; the kinematics fix the
lepton energy to be $z = 1 + \varepsilon$ where
\begin{equation}
\varepsilon := \frac{m_\ell^2}{\Mbc^2}
\end{equation}
is the rescaled lepton mass squared. Clearly $\varepsilon$ is a small quantity
for the electron and the muon case, in which we can use only the leading
term. However, for the $\tau$ the parameter $\varepsilon$ is sizable and we
thus keep the full dependence on $\varepsilon$.

The coefficient ${\cal A}$ contains contributions from both the spin singlet
and the spin triplet dimension-6 operator:
\begin{equation}
{\cal A} = {\cal A}_{({\rm 1} \times {\rm 1})}[{}^1\!S_0^{(1)}]
		\bra{B_c} {\cal O}_{(1 \times 1)}[{}^1\!S_0^{(1)}] \ket{B_c}
	 + {\cal A}_{({\rm 1} \times {\rm 1})}[{}^3\!S_1^{(1)}]
		\bra{B_c} {\cal O}_{(1 \times 1)}[{}^3\!S_1^{(1)}] \ket{B_c}\,.
\end{equation}
We expect that the leading term represented by ${\cal A}_{({\rm 1} \times {\rm
1})}[{}^1\!S_0^{(1)}]$ is suffering from helicity suppression just like the
purely leptonic decay. In the inclusive case this is due to spin symmetry
making its contribution to $R_{\mu \nu}$ entirely proportional to $v_\mu
v_\nu$. If vacuum insertion is applied, this contribution becomes exactly equal
to the one from the purely leptonic case. In fact $\Gamma_{\!0} {\cal A}_{({\rm
1} \times {\rm 1})}[{}^1\!S_0^{(1)}]$ reproduces the total rate (\ref{Gamlept})
of the purely leptonic decay: Defining
\begin{equation}
\Gamma_{\!0} := \frac{G_F^2}{8\pi M} |V_{cb}|^2 \Mbc^2
\end{equation}
the short distance coefficient reads
\begin{equation}
\label{A1S0}
{\cal A}_{({\rm 1} \times {\rm 1})}[{}^1\!S_0^{(1)}]
= \varepsilon (1 - \varepsilon)^2 \,.
\end{equation}

While the Wilson coefficient of the spin singlet operator explicitly shows a
factor $\varepsilon$ we do not expect any helicity suppression in the spin
triplet case: We obtain for the corresponding short distance
coefficient
\begin{equation}
\label{A3S1}
{\cal A}_{({\rm 1} \times {\rm 1})}[{}^3\!S_1^{(1)}]
=  \frac{(1 - \varepsilon)^2 (2 + \varepsilon)}{3}
\end{equation}
which does not vanish as $\varepsilon \to 0$. From this observation we
deduce that the spin triplet operator could become important even
though its NRQCD matrix element is suppressed by $\lambda^2 \bar{v}^6$ with
respect to the leading matrix element of the operator ${\cal O}_{(1 \times
1)}[{}^1\!S_0^{(1)}]$. If we compare the magnitude of the suppression factors
\begin{equation}
\label{weight}
\frac{{\cal A}_{(1 \times 1)}[{}^1\!S_0^{(1)}]}
     {{\cal A}_{(1 \times 1)}[{}^3\!S_1^{(1)}]} \approx
\frac{3}{2} \left( \frac{m_\ell}{\Mbc} \right)^2
\quad \leftrightarrow \quad \lambda^2 \bar{v}^6 \approx
\frac{\bra{B_c} {\cal O}_{(1 \times 1)}[{}^3\!S_1^{(1)}] \ket{B_c}}
     {\bra{B_c} {\cal O}_{(1 \times 1)}[{}^1\!S_0^{(1)}] \ket{B_c}}
\end{equation}
we realize that the spin triplet contribution overcomes the spin singlet
contribution if the lepton mass is small enough. For sure this condition is
fulfilled in the electron case. For muons both contributions are comparable
although it depends on the value of $\lambda$ and $\bar{v}$ for $B_c$. In the
case $\ell = \tau$ we can neglect the spin triplet part.

The coefficients ${\cal B}_i$ arise with a factor $1/(2m)^2$ where $m$ is the
effective (= twice the reduced) mass of the $b$ quark and the $c$ antiquark.
Furthermore we have introduced the dimensionless variable $\eta$ to
parameterize their mass difference:
\begin{equation}
m := \frac{2 m_b m_c}{m_b + m_c} \,, \qquad
\eta := \frac{m_b - m_c}{m_b + m_c} \,.
\end{equation}
Generally operators of higher powers in the inverse mass $m$ are accounted to
be of higher order. However, there is a big {\it caveat} in the endpoint region
of the energy spectrum. Because the expansion is an expansion in terms of
$1/(\Mbc - 2E_\ell)$ rather than in terms of $1/(2m)$ where
\begin{equation}
\label{massrel}
2m = (1 - \eta^2) \Mbc
\end{equation}
it breaks down in the endpoint region. In the framework of NRQCD it was shown
that the availability of the theoretical prediction can be extended to higher
energy values if the contributions of the order $[\mQ \bar{v}^2/(\mQ -
E_\ell)]^n$ are summed up to so-called shape functions \cite{Rothstein:1997ac}.
Thereby the endpoint also shifts from the partonic value characterized by mass
of the $\QQ$ pair to the hadronic value which is a function of the mass of the
decaying particle. In our case higher order derivatives of the leading delta
function $\delta(1 + \varepsilon - z)$ represent contributions to this shape
function. We will elaborate this in Subsection \ref{shape} more detailed.

First we present the result for the coefficients ${\cal B}_i$ in (\ref{dGdz}).
Due to heavy quark spin symmetry the spin triplet operators get an additional
suppression factor compared to the spin singlet operators when they are
inserted into the $B_c$ matrix element. On the other hand they should not
suffer from helicity suppression as we have seen from ${\cal O}_{({\rm 1}
\times {\rm 1})}[{}^3\!S_1^{(1)}]$. Therefore we will not only keep the spin
singlet contributions to ${\cal B}_0$ but also the most relevant triplet
operator ${\cal O}_{({\rm R} \times {\rm R})}[{}^3\!P_J^{(1)}]$ according to
Fig.~\ref{fig1}:
\begin{equation}
\begin{split}
{\cal B}_0 & =
{\cal B}^{(0)}_{({\rm C} \times {\rm C})}[{}^1\!S_0^{(1)}]
\bra{B_c} {\cal O}_{({\rm C} \times {\rm C})}[{}^1\!S_0^{(1)}] \ket{B_c}
+ {\cal B}^{(0)}_{({\rm C} \times {\rm R})}[{}^1\!S_0^{(1)}]
\bra{B_c} {\cal O}_{({\rm C} \times {\rm R})}[{}^1\!S_0^{(1)}] \ket{B_c}
\\
& \text{}
+ {\cal C}^{(0)}_{({\rm R} \times {\rm R})}[{}^1\!S_0^{(1)}]
\bra{B_c} {\cal P}_{({\rm R} \times {\rm R})}[{}^1\!S_0^{(1)}] \ket{B_c}
+ {\cal B}^{(0)}_{({\rm R} \times {\rm R})}[{}^1\!P_1^{(1)}]
\bra{B_c} {\cal O}_{({\rm R} \times {\rm R})}[{}^1\!P_1^{(1)}] \ket{B_c}
\\
& \text{}
+ {\cal B}^{(0)}_{({\rm R} \times {\rm R})}[{}^3\!P_0^{(1)}]
\bra{B_c} {\cal O}_{({\rm R} \times {\rm R})}[{}^3\!P_0^{(1)}] \ket{B_c} \,.
\end{split}
\end{equation}

For light leptons the pieces which do no suffer from helicity suppression can
be as sizable as the dimension-6 contribution ${\cal A}$. If we consider the
limit $\varepsilon \to 0$ approximately fulfilled by electrons we should find
non-vanishing contributions from ${\cal B}^{(0)}_{({\rm R} \times {\rm
R})}[{}^1\!P_1^{(1)}]$ and ${\cal B}^{(0)}_{({\rm R} \times {\rm
R})}[{}^3\!P_0^{(1)}]$ where the quark-antiquark pair is in a $P$-wave state.
Additionally we expect the coefficients ${\cal B}^{(0)}_{({\rm C} \times {\rm
C})}[{}^1\!S_0^{(1)}]$ and ${\cal B}^{(0)}_{({\rm C} \times {\rm
R})}[{}^1\!S_0^{(1)}]$ to survive in this limit. There the soft degrees of
freedom moving relative to the quark-antiquark pair carry away some spin or
angular momentum. Indeed ${\cal C}^{(0)}_{({\rm R} \times {\rm
R})}[{}^1\!S_0^{(1)}]$ is the only piece of ${\cal B}_0$ that is proportional
to $\varepsilon$:
\begin{subequations}
\begin{align}
{\cal B}^{(0)}_{({\rm C} \times {\rm C})}[{}^1\!S_0^{(1)}]
&= \frac{2 (6 - 8\eta^2 + 3\eta^4)}{3} - \varepsilon \left[
1 + \eta^2 + 2 (3 - 6\eta^2 + 2 \eta^4) \varepsilon
		- \frac{3 - 5\eta^2}{3} \varepsilon^2 \right] ,
\\
{\cal B}^{(0)}_{({\rm C} \times {\rm R})}[{}^1\!S_0^{(1)}]
&= \frac{8 \eta (2 - \eta^2)}{3} - 8 \eta \varepsilon \left[
1 - \varepsilon + \frac{2 - \eta^2}{3} \varepsilon^2 \right] ,
\\
{\cal C}^{(0)}_{({\rm R} \times {\rm R})}[{}^1\!S_0^{(1)}]
&= - 4 \varepsilon \left[
1 - 2 (2 - \eta^2) \varepsilon
+ (3 - 2\eta^2) \varepsilon^2 \right] ,
\\
{\cal B}^{(0)}_{({\rm R} \times {\rm R})}[{}^1\!P_1^{(1)}]
&= \frac{8 \eta^2}{3} 
- \frac{4 \eta^2}{3} \varepsilon \left[ 3 - \varepsilon^2 \right] ,
\\
{\cal B}^{(0)}_{({\rm R} \times {\rm R})}[{}^3\!P_0^{(1)}]
&= 8 - 4 \varepsilon \left[
3 (1 - \eta^2) + 6 \eta^2 \varepsilon - (1 + 3 \eta^2) \varepsilon^2 \right] .
\end{align}
\end{subequations}
Note that the terms odd in $\eta$ changes their sign if we consider $B_c^+ \to
X \ell^+ \nu_\ell$ instead of $B_c^- \to X \ell \bar{\nu}_\ell$. However, this
does not mean that the result violates $CP$ invariance, since the sign of the
operator ${\cal O}_{({\rm C} \times {\rm R})}[{}^1\!S_0^{(C)}]$ also changes.
Furthermore we have used heavy quark spin symmetry to summarize the $P$ wave
channels ${}^3\!P_0$, ${}^3\!P_1$, and ${}^3\!P_2$:
\begin{equation}
\bra{B_c} {\cal O}_{({\rm R} \times {\rm R})}[{}^3\!P_J^{(1)}] \ket{B_c}
= (2J + 1)
\bra{B_c} {\cal O}_{({\rm R} \times {\rm R})}[{}^3\!P_0^{(1)}] \ket{B_c} \,.
\end{equation}

Next we deal with the contributions ${\cal B}_1$ and ${\cal B}_2$. As mentioned
above they become relevant close to the endpoint since they are more singular
than the leading contribution. Once the $\delta$ function and its derivatives
are resummed we will find a shape function enhancement close to the endpoint.
For reasons which will become clear in a moment we also keep the ${\cal B}_1$
pieces stemming from ${\cal O}_{({\rm C} \times {\rm C})}[{}^3\!S_1^{(1)}]$ and
${\cal P}_{({\rm C} \times {\rm R})}[{}^3\!S_1^{(1)}]$:
\begin{equation}
\begin{split}
{\cal B}_1 & =
{\cal B}^{(1)}_{({\rm C} \times {\rm C})}[{}^1\!S_0^{(1)}]
\bra{B_c} {\cal O}_{({\rm C} \times {\rm C})}[{}^1\!S_0^{(1)}] \ket{B_c}
+ {\cal B}^{(1)}_{({\rm C} \times {\rm R})}[{}^1\!S_0^{(1)}]
\bra{B_c} {\cal O}_{({\rm C} \times {\rm R})}[{}^1\!S_0^{(1)}] \ket{B_c}
\\
& \text{}
+ {\cal C}^{(1)}_{({\rm R} \times {\rm R})}[{}^1\!S_0^{(1)}]
\bra{B_c} {\cal P}_{({\rm R} \times {\rm R})}[{}^1\!S_0^{(1)}] \ket{B_c}
+ {\cal B}^{(1)}_{({\rm C} \times {\rm C})}[{}^3\!S_1^{(1)}]
\bra{B_c} {\cal O}_{({\rm C} \times {\rm C})}[{}^3\!S_1^{(1)}] \ket{B_c}
\\
& \text{}
+ {\cal B}^{(1)}_{({\rm C} \times {\rm R})}[{}^3\!S_1^{(1)}]
\bra{B_c} {\cal O}_{({\rm C} \times {\rm R})}[{}^3\!S_1^{(1)}] \ket{B_c}
+ {\cal C}^{(1)}_{({\rm R} \times {\rm R})}[{}^3\!S_1^{(1)}]
\bra{B_c} {\cal P}_{({\rm R} \times {\rm R})}[{}^3\!S_1^{(1)}] \ket{B_c} \,.
\end{split}
\end{equation}

While the spin singlet contributions are still helicity suppressed the
coefficients corresponding to a quark-antiquark pair in a spin triplet state do
not vanish for $\varepsilon \to 0$:
\begin{subequations}
\begin{align}
{\cal B}^{(1)}_{({\rm C} \times {\rm C})}[{}^1\!S_0^{(1)}]
&= \frac{1 - \eta^2}{6} \varepsilon \left[
13 - 12\eta^2 - 3 (7 - 6\eta^2) \varepsilon + 3 \varepsilon^2
- (11 - 10\eta^2) \varepsilon^3 \right] ,
\\
{\cal B}^{(1)}_{({\rm C} \times {\rm R})}[{}^1\!S_0^{(1)}]
&= - \frac{2 \eta (1 - \eta^2)}{3} \varepsilon \left[
1 - \varepsilon \right]^3 ,
\\
{\cal C}^{(1)}_{({\rm R} \times {\rm R})}[{}^1\!S_0^{(1)}]
&= - 2 (1 - \eta^2) \varepsilon \left[ 1 - \varepsilon \right]^3 ,
\\
\begin{split}
\label{B3S1cc1}
{\cal B}^{(1)}_{({\rm C} \times {\rm C})}[{}^3\!S_1^{(1)}]
&= - \frac{1 - \eta^2}{3} + \frac{1 - \eta^2}{30} \varepsilon \big[
29 + \eta^2 - 15 (1 + \eta^2) \varepsilon
\\
& \hspace{10em} \text{} - (17 - 27\eta^2) \varepsilon^2
- 15 (1 - \eta^2) \varepsilon^3 \big] \,,
\end{split}
\\
\label{B3S1cr1}
{\cal B}^{(1)}_{({\rm C} \times {\rm R})}[{}^3\!S_1^{(1)}]
&= - \frac{4 \eta (1 - \eta^2)}{3} + \frac{\eta (1 - \eta^2)}{3} \varepsilon
\left[ 11 - 9\varepsilon + \varepsilon^2 + \varepsilon^3 \right] ,
\\
\label{C3S1rr1}
{\cal C}^{(1)}_{({\rm R} \times {\rm R})}[{}^3\!S_1^{(1)}]
&= - \frac{4 (1 - \eta^2)}{3} + \frac{2 (1 - \eta^2)}{3} \varepsilon \left[
5 - 3\varepsilon - \varepsilon^2 + \varepsilon^3 \right] .
\end{align}
\end{subequations}

Finally we quote the results for the coefficient ${\cal B}^{(2)}$. In that case
we only find contributions to RCM $\times$ RCM operators:
\begin{equation}
\begin{split}
{\cal B}_2 & =
{\cal B}^{(2)}_{({\rm C} \times {\rm C})}[{}^1\!S_0^{(1)}]
\bra{B_c} {\cal O}_{({\rm C} \times {\rm C})}[{}^1\!S_0^{(1)}] \ket{B_c}
+ {\cal B}^{(2)}_{({\rm C} \times {\rm C})}[{}^3\!S_1^{(1)}]
\bra{B_c} {\cal O}_{({\rm C} \times {\rm C})}[{}^3\!S_1^{(1)}] \ket{B_c}
\\
& \text{}
+ {\cal C}^{(2)}_{({\rm C} \times {\rm C})}[{}^3\!S_1^{(1)}]
\bra{B_c} {\cal P}_{({\rm C} \times {\rm C})}[{}^3\!S_1^{(1)}] \ket{B_c}
\end{split}
\end{equation}
where 
\begin{subequations}
\begin{align}
{\cal B}^{(2)}_{({\rm C} \times {\rm C})}[{}^1\!S_0^{(1)}]
&= \frac{(1 - \eta^2)^2}{6} \varepsilon \left[ 1 - \varepsilon^2 \right]^2 ,
\\
\label{B3S1cc2}
{\cal B}^{(2)}_{({\rm C} \times {\rm C})}[{}^3\!S_1^{(1)}]
&= \frac{7 (1 - \eta^2)^2}{60} - \frac{(1 - \eta^2)^2}{60} \varepsilon \left[
25 - 44 \varepsilon + 38 \varepsilon^2 - 9 \varepsilon^3 - 3 \varepsilon^4
\right] ,
\\
\label{C3S1cc2}
{\cal C}^{(2)}_{({\rm C} \times {\rm C})}[{}^3\!S_1^{(1)}]
&= - \frac{(1 - \eta^2)^2}{15} + \frac{(1 - \eta^2)^2}{15} \varepsilon \left[
5 - 12 \varepsilon + 14 \varepsilon^2 - 7 \varepsilon^3 + \varepsilon^4
\right] .
\end{align}
\end{subequations}

Again only the spin singlet pieces show helicity suppression. However, this
feature of the coefficients of higher order derivatives of the delta function
does not surprise since it is given by the phase space structure of the
process as we will explain in the following subsection.

\subsection{Shape function formalism}
\label{shape}

The spectrum within the operator product expansion is given as a series of
$\delta$ functions and its derivatives and thus is not physical. However, it
can be interpreted within the shape function formalism. As mentioned before the
NRQCD expansion breaks down for lepton energies to close to the endpoint. The
physical reason for this breakdown comes from the lack of phase space for
radiating off soft gluons if the lepton energy reaches it maximum value. Hence
in the endpoint region the non-perturbative subprocess $B_c \to (\bar{b}c) +
\text{\it soft degrees of freedom}$ has to be parametrized by so-called shape
functions reflecting the phase space dependence rather than just by numbers
like the NRQCD matrix elements.

The formal construction of the shape functions in $B_c$ decays proceeds
analogously to the one in quarkonia decays \cite{Rothstein:1997ac}. The phase
space of the short distance subprocess is given by
\begin{equation}
d{\rm PS} = \widetilde{dp_\ell} \widetilde{dp_{\bar{\nu}_\ell}}
  (2 \pi)^4 \delta^4(P + l - p_\ell - p_{\bar{\nu}_\ell})
\end{equation}
where $P = \Mbc v$ and $l$ are the main and the residual cms momenta of the
heavy quark-antiquark system in the meson rest frame, respectively. If we
integrate over $d^4p_{\bar{\nu}_\ell}$ and expand the lepton energy spectrum in
the small residual cms momentum we recognize that only the light cone component
$l_+ = \hat{n} \cdot l$ is relevant for the soft gluon dynamics. Here the light
cone vector is given by $\hat{n} = 2(P - p_\ell)/\Mbc$. Then the lepton energy
spectrum reads\footnote{We ignore the lepton mass $m_\ell$ in this discussion.}
\begin{equation}
\label{dGdEshape}
\frac{d\Gamma}{dE_\ell}
= \sum_n {\cal C}[n] \int \!\! dl_+ \,
f_{B_c}[n](l_+) \, \delta(\Mbc/2 - E_\ell + l_+/2)
\end{equation}
where the shape functions are defined via
\begin{equation}
\label{shapefunc}
f_{B_c}[n](l_+)
= \bra{B_c} \left[ \psi^\dag {\cal K}_n \chi \right]
\delta(l_+ - iD_+) \left[ \chi^\dag {\cal K}'_n \psi \right] \ket{B_c} \,.
\end{equation}

As for the matrix elements there is one shape function per each ($b\bar{c}$)
configuration indicated by the kernels ${\cal K}_n$ and ${\cal K}'_n$ in
(\ref{shapefunc}). In case of spin singlet configurations vacuum saturation is
applicable and the covariant derivative $iD_+$ reduces to the time component
which picks up the binding energy $M - \Mbc$ of the $B_c$. Thus the shape
function shifts the endpoint of the lepton energy spectrum from the unphysical
partonic value $\Mbc/2$ to the physical $M/2$ expressed by the hadronic $B_c$
mass.

Since the shape function resums purely kinematical effects originating from the
phase space the short distance coefficients ${\cal C}[n]$ in
eq.~(\ref{dGdEshape}) are not affected at all. This means particularly that the
coefficient (\ref{A1S0}) of the shape function $f_{B_c}[{}^1\!S_0^{(1)}]$ is
still helicity suppressed while the coefficient (\ref{A3S1}) of the spin
triplet shape function $f_{B_c}[{}^3\!S_1^{(1)}]$ do not vanish for
$\varepsilon \to 0$.

After changing to our four-dimensional notation and expanding the shape
functions in the residual momentum $iD_+/\Mbc$ we reobtain the spectrum in
terms of delta functions and their derivatives like eq.~(\ref{dGdz}). Acting
on the spin triplet structure of the dimension-6 operator ${\cal O}_{(1 \times
1)}[{}^3\!S_1^{(1)}]$ the zeroth component of the covariant derivative
generates the following dimension-8 operators (cf.~eq.~(\ref{massrel})):
\begin{equation}
\begin{split}
& \frac{1}{2} \left\{
\bigg[ \bar{h}_v^{(+)} \, \gamma^\perp_\mu \, h_v^{(-)} \bigg] \left[
\frac{(ivD)}{\Mbc} \! \left( \bar{h}_v^{(-)} \, \gamma_\perp^\mu \, h_v^{(+)}
\right) \right] + {\rm h.c.} \right\}
\\
& \hspace{3em} = \frac{1}{4m^2} \left( - \frac{1 - \eta^2}{2} \right) \left\{
{\cal O}_{({\rm C} \times {\rm C})}[{}^3\!S_1^{(1)}]
+ 4 \eta {\cal O}_{({\rm C} \times {\rm R})}[{}^3\!S_1^{(1)}]
+ 4 {\cal P}_{({\rm R} \times {\rm R})}[{}^3\!S_1^{(1)}]
\right\}
\end{split}
\end{equation}

We re-observe the same structure as we have seen in the pieces of the
coefficients (\ref{B3S1cc1}-\ref{C3S1rr1}) that remain in the limit
$\varepsilon \to 0$. Actually these pieces are completely determined by the
coefficient (\ref{A3S1}) of ${\cal O}_{(1 \times 1)}[{}^3\!S_1^{(1)}]$ and the
expansion of the corresponding shape function. Similarly one can derive the
non-vanishing pieces in (\ref{B3S1cc2}) and (\ref{C3S1cc2}) by applying
$iD_\perp/\Mbc$ two times on ${\cal O}_{(1 \times 1)}[{}^3\!S_1^{(1)}]$.

Note that the shape function do not determine the complete coefficients ${\cal
B}_1$ and ${\cal B}_2$. The pieces proportional to $\varepsilon$ contain terms
which cannot be directly derived from the shape function. Hence the structure
of the spin singlet coefficients in ${\cal B}_1$ and ${\cal B}_2$ is more
complicated. Furthermore we want to stress that the origin of the contributions
to ${\cal B}_0$ which do not vanish in the limit of massless leptons is not
kinematical. To explain these pieces it is not sufficient to introduce shape
functions. Rather one has to consider terms that are generated dynamically by
the non-perturbative physics inside the $B_c$ taken into account in the
bilinear expansion (\ref{bilin}).

\subsection{Moments of the spectrum}

Another way to understand the unphysical series of delta functions and their
derivatives is by the means of moments which implies some smearing
\cite{Poggio:1976af}.

The two-body structure of the decay at tree level yields a spectrum in terms of
an expansion in singular functions
\begin{equation} \label{sing}
\frac{d\Gamma}{dz}
= \sum_{k=0}^\infty \frac{M_k}{k!} \, \delta^{(k)}(1 + \varepsilon - z)
\end{equation}
where the coefficients of the expansion are given by the moments
\begin{equation}
M_k = \int\limits_0^{1+\varepsilon} \!\! dz \,
\frac{d\Gamma}{dz} (1 + \varepsilon - z)^k \,.
\end{equation}

A comparison of the result (\ref{dGdz}) with data can thus be performed by
sampling the data into moments, for which we get a theoretical prediction. The
zeroth moment $M_0$ is simply the total rate for which we get
\begin{equation}
\Gamma = \Gamma_{\!0} \left[ {\cal A} + \frac{1}{(2m)^2} \, {\cal B}_0 \right]
\end{equation}
while the first an the second moments are at least of order $1/m^2$
\begin{equation}
M_1 = \frac{\Gamma_{\!0}}{(2m)^2} \, {\cal B}_1 \,,
\qquad M_2 = 2 \frac{\Gamma_{\!0}}{(2m)^2} \, {\cal B}_2 \,.
\end{equation}

Hence measuring the second moments of the lepton energy spectra in the
semi-leptonic $B_c$ decay provides direct access on the order of magnitude of
the RCM $\times$ RCM sector. The fact that the values for the matrix elements
do not depend on the type of the lepton in the final state may help to improve
the accuracy of this determination. Likewise measurements of the first and
zeroth moments may provide some information for the matrix elements of the
other operators.

\subsection{Total rate in the limit $m_\ell \to 0$}

Finally we discuss the limit of vanishing lepton mass. In this limit only
subleading terms in $\bar{v}$ survive since the leading term has to vanish due
to helicity arguments. On the other hand, this limit is certainly a good
approximation in the case of the electron spectrum.

For $\varepsilon \to 0$ non-vanishing contribution to the lepton energy
spectrum (\ref{dGdz}) contains a piece which is proportional to $\delta(1 -
z)$:
\begin{equation}
\label{ml20}
\begin{split}
\frac{d\Gamma}{dz} & = \Gamma_{\!0} \bigg\{ \bigg(
\frac{2}{3} \bra{B_c} {\cal O}_{(1 \times 1)}[{}^3\!S_1^{(1)}] \ket{B_c}
+ \frac{1}{6m^2} \Big[
  (6 - 8\eta^2 + 3\eta^4)
           \braket{{\cal O}_{({\rm C} \times {\rm C})}[{}^1\!S_0^{(1)}]}
\\
&\text{} \hspace{2em} + 4 \eta (2 - \eta^2)
           \braket{{\cal O}_{({\rm C} \times {\rm R})}[{}^1\!S_0^{(1)}]}
+ 4 \eta^2 \braket{{\cal O}_{({\rm R} \times {\rm R})}[{}^1\!P_1^{(1)}]}
+ 12       \braket{{\cal O}_{({\rm R} \times {\rm R})}[{}^3\!P_0^{(1)}]}
\Big] \bigg) \delta(1 - z)
\\
&\text{} \hspace{2em} - \frac{1 - \eta^2}{12m^2} \left[
         \braket{{\cal O}_{({\rm C} \times {\rm C})}[{}^3\!S_1^{(1)}]}
+ 4 \eta \braket{{\cal O}_{({\rm C} \times {\rm R})}[{}^3\!S_1^{(1)}]}
+ 4      \braket{{\cal P}_{({\rm R} \times {\rm R})}[{}^3\!S_1^{(1)}]}
\right] \delta'(1 - z)
\\
&\text{} \hspace{2em} + \frac{(1 - \eta^2)^2}{60m^2} \left[
  7 \braket{{\cal O}_{({\rm C} \times {\rm C})}[{}^3\!S_1^{(1)}]}
- 4 \braket{{\cal P}_{({\rm C} \times {\rm C})}[{}^3\!S_1^{(1)}]}
\right] \delta''(1 - z) \bigg\}
\end{split}
\end{equation}
This piece will contribute to the total rate of the semi-leptonic decay $B_c
\to X \ell \bar{\nu}_\ell$ even if the energy and momentum of the light hadrons
in the final state are extremely soft. Independent of the exact values of the
counting parameters $\kappa$, $\lambda$, and $\bar{v}$ this piece should be
sizable whereas the purely leptonic $B_c$ decay rate ($\ref{Gamlept}$) vanishes
for $m_\ell \to 0$. Consequently we conclude that the semi-leptonic rate
exceeds the leptonic one by orders of magnitude for $\ell = e$ while it at
least reaches it in the muon case (cf.~eq.~\ref{weight}). Since we expect the
light hadrons in the final state to be rather soft the experimental search for
the exclusive channel $B_c \to \ell \bar{\nu}_\ell$ become very difficult
unless the matrix elements in (\ref{ml20}) are unnaturally small.

\section{Conclusions}

We have investigated the semi-leptonic decay $B_c \to X \ell \bar{\nu}_\ell$
where $X$ denotes light hadrons below the $D^0$ threshold. This decay provides
a unique testing ground in quarkonia physics, since the two heavy quarks have
to annihilate through weak interaction and hence one may investigate in this
way the light degrees of freedom in a quarkonium.  The total semi-leptonic rate
into light hadrons does not show any helicity suppression at subleading order
and thus can exceed the purely leptonic one for light leptons, i.e.~for
electrons and possibly also for muons. Thus the measurement of the leptonic
$B_c$ decay rate could become a quite complicated maybe even infeasible
experimental task.

The $B_c$ is an intermediate case between the $\Upsilon$ and the $J/\psi$.
While NRQCD is believed to be a reasonable approximation at least for the
low-lying states of the $\Upsilon$, the non-relativistic approximation becomes
questionable for the $J/\psi$. This point is of relevance for power counting,
since the usual NRQCD power counting may not be valid for the $J/\psi$.  This
will also affect the contributions related to the centre-of-mass movement of
the heavy quark-antiquark pair inside a quarkonia-like bound state. For this
reason we propose a slightly generalized power counting scheme with respect to
standard NRQCD velocity rules taking care of possible contributions from the
cms-motion.

We have shown that the leptonic energy spectrum provides a testing ground for
estimating the relevance of such contributions. Measurements of the moments of
the spectrum may also help to clarify the power counting rules for spin-flip
transitions which are a recent subject of investigation in charmonia physics.

\section*{Acknowledgements}

We thank Andreas Schenk for his collaboration during earlier stages of this
work. Furthermore we want to thank Martin Beneke for enlighting discussions on
the question of power counting. S.W.~acknowledges useful discussions with
M.A.~Sanchis Lozano and the support by a research grant of the Deutsche
Forschungsgemeinschaft (DFG). T.M.~thanks the Aspen Center for Physics for its
hospitality during the summer workshop and acknowledges support from the
Bundesinisterium f\"ur Bildung und Forschung (bmb+f) and from the DFG
Forschergruppe ``Quantenfeldtheorie, Computeralgebra und Monte Carlo
Simulationen''.

\begin{appendix}

\section{Bilinears}

To expand the currents $\bar{Q}_v^{(\pm)} \, \Gamma \, Q_v^{(\mp)}$
we use the field expansion that can be read off if the effective Lagrangian
(\ref{L_eff}) is compared with the full QCD Lagrangian. Additionally
considering eq.~(\ref{iDiD}) and the equations of motion (\ref{eom}) this
yields
\begin{equation}
Q_v^{(\pm)}(x)
= \left( 1 + \frac{1}{2m_\pm} \, i\fmslash{D}_\perp
           + \frac{1}{8m_\pm^2} \left(iD_\perp \right)^2 \right) h_v^{(\pm)}(x)
+ {\cal O}\left(\frac{1}{m_\pm^3}\right) .
\end{equation}

Instead of using the masses of the quark ($m_+$) and the antiquark ($m_-$) we
present the expansion in terms of the effective mass $m$ and the dimensionless
parameter $\eta$ parameterizing the mass difference:
\begin{equation}
m := \frac{2 m_+ m_-}{m_+ + m_-} \qquad \text{and} \qquad
\eta := \frac{m_+ - m_-}{m_+ + m_-}
\end{equation}
Then the bilinear expansion up to ${\cal O}(1/m^2)$ reads:
\begin{equation}
\label{bilin}
\begin{split}
\bar{Q}_v^{(+)} \, \Gamma \, Q_v^{(-)}
&= \bar{h}_v^{(+)} \, \Gamma \, h_v^{(-)}
\\
&+ \frac{1}{4m} \, \left\{ \;
       i\partial_\perp^\mu \left(
          \bar{h}_v^{(+)} \, [ \Gamma, \gamma_\mu ] \, h_v^{(-)}
       \right)
     + \bar{h}_v^{(+)} \, \{ \Gamma, \iDslr_\perp \} \, h_v^{(-)}
\right.
\\
& \left. \hspace{1.4em} \text{}
     + \eta \left[
          i\partial_\perp^\mu \left(
             \bar{h}_v^{(+)} \, \{ \Gamma, \gamma_\mu \} \, h_v^{(-)}
          \right)
        + \bar{h}_v^{(+)} \, [ \Gamma, \iDslr_\perp ] \, h_v^{(-)}
   \right]
\right\}
\\
&+ \left( \frac{1}{4m} \right)^2 \Bigg\{
  (1 + \eta^2) \left[
     (i\partial_\perp)^2 \left(
        \bar{h}_v^{(+)} \, \Gamma \, h_v^{(-)}
     \right)
   + \bar{h}_v^{(+)} \, \Gamma (\iDlr_\perp)^2 \, h_v^{(-)}
  \right]
\\
& \hspace{4.4em} \text{}
- (1 - \eta^2) \, \Bigg[
     i\partial_\perp^\mu i\partial_\perp^\nu \left(
        \bar{h}_v^{(+)} \, \frac{1}{2} \{
           \gamma_\mu \Gamma \gamma_\nu + \gamma_\nu \Gamma \gamma_\mu
        \} \, h_v^{(-)}
     \right)
\\
& \hspace{9.2em} \text{}
  - \bar{h}_v^{(+)}\frac{1}{2}\{
           \gamma_\mu \Gamma \gamma_\nu + \gamma_\nu \Gamma \gamma_\mu
        \} \iDlr_\perp^\mu \iDlr_\perp^\nu \, h_v^{(-)}
  \Bigg]
\\
& \hspace{4.4em} \text{}
+ 4 \, \eta \; i\partial_\perp^\mu \left(
                   \bar{h}_v^{(+)} \, \Gamma \iDlr^\perp_\mu \, h_v^{(-)}
  \right)
\Bigg\} + {\cal O}\left(\frac{1}{m^3}\right) \,.
\end{split}
\end{equation}
In the case of colour octet case ($\bar{Q}_v^{(+)} \, \Gamma \, T^a \,
Q_v^{(-)}$):
\begin{itemize}
\item insert $T^a$ into the currents $\bar{h}_v^{(+)} \, \ldots \, h_v^{(-)}$
\item replace $i\partial_\perp^\mu$ by the covariant derivative $iD_\perp^\mu$
\end{itemize}
Finally the corresponding formula for $\bar{Q}_v^{(-)} \, \Gamma C \,
Q_v^{(+)}$ is obtained by the replacement $\eta \to \text{} - \eta$.

\end{appendix}

\end{document}